\documentclass[aps,showpacs,twocolumn,nofootinbib]{revtex4}
\usepackage{graphicx}
\usepackage{epstopdf}
\usepackage{dcolumn}
\usepackage{amsmath}

\begin{document}
\title{Properties of isoscalar-pair condensates}
\author{P.~Van~Isacker}
\affiliation{Grand Acc\'el\'erateur National d'Ions Lourds,
CEA/DRF--CNRS/IN2P3, Bvd Henri Becquerel, F-14076 Caen, France}

\author{A.~O.~Macchiavelli and P.~Fallon}
\affiliation{Nuclear Science Division, Lawrence Berkeley National Laboratory,
Berkeley, California 94720}

\author{S.~Zerguine}
\affiliation{Department of Physics, PRIMALAB Laboratory,
University of Batna, Avenue Boukhelouf M El Hadi, 05000 Batna, Algeria}

\date{\today}

\begin{abstract}
It is pointed out that the ground state
of $n$ neutrons and $n$ protons in a single-$j$ shell,
interacting through an isoscalar ($T=0$) pairing force,
is not paired, $J=0$, but rather spin-aligned, $J=n$.
This observation is explained in the context of a model of isoscalar $P$ ($J=1$) pairs,
which is mapped onto a system of $p$ bosons,
leading to an approximate analytic solution
of the isoscalar-pairing limit in $jj$ coupling.
\end{abstract}

\maketitle

\section{Introduction}
\label{s_intro}
In 1958, Bohr, Mottelson, and Pines~\cite{bmp58} suggested a possible analogy
between the excitation spectra of nuclei and those of the superconducting metallic state.
Since then, a wealth of experimental data have been accumulated,
supporting the important role played by pairing correlations
in defining properties of atomic nuclei,
such as deformation, moments of inertia, alignments, etc.~\cite{Brink05,BCS50}.
Today, the study of pairing correlations
continues to be a subject of active research in nuclear physics,
with an emphasis in exotic nuclei.
Of particular interest is the understanding of the role
played by the isoscalar $(T=0)$ and isovector $(T=1)$ pairing forces~\cite{nprev}
in the structure of  $N\approx Z$ nuclei.

Given the charge independence of the nuclear force,
$T=1$ pairing is on an equal footing
between the $T_z=0$ neutron-proton (np)
and  $|T_z|=1$ neutron-neutron and proton-proton (nn and pp) components.
In addition, we have the unique possibility
of studying the formation of a condensate of $T=0$ np pairs, 
thus implying the possible co-existence of ``Cooper" pairs of isoscalar and isovector type.    
Although the nuclear force is stronger in the $T=0$ channel,
it is still not clear how effective the ({\it in-medium}) $T=0$ correlations are
in giving rise to a ground-state isoscalar condensate~\cite{nprev}.

In this paper we consider some interesting properties of the isoscalar condensate in the $jj$ coupling scheme,
in particular with regards to its angular momentum.  
Our motivation starts by  studying the numerical results of a shell-model calculation,
within the space of single-particle spin-orbit partners,
showing that when the isoscalar component is dominant,
the ground-state is not paired to $J^\pi=0^+$
but, rather, it behaves as a state of aligned $1^+$ {\it quasi-deuterons}.
To gain further insight into the peculiar structure of these condensates,
we develop a boson mapping of the shell model,
leading to an approximate analytic solution.
Group-theoretical solutions of the pairing problem
are known in the isoscalar and isovector limits of $LS$ coupling~\cite{Flowers64,Pang69,Evans81}
and in the isovector limit of $jj$ coupling~\cite{Racah52,Kerman61}
but, to our knowledge, not in the isoscalar limit of $jj$ coupling.
Our results therefore provide, for the first time, approximate analytic formulae
for the energies of the lowest states in that case.
While we are of course aware that this limit is not applicable to real nuclei,
these states might exist close to the ground state
in specific regions of the $N=Z$ line~\cite{Gezerlis11,Bulthuis16}
and, more interestingly perhaps, could be realized in atomic traps.

\section{Single-$j$ shell model}
\label{s_sm}
Single-shell models that capture the main ingredients of the problem
provide a useful framework to understand
the competition of isovector and isoscalar pairing interactions.
Here we start by considering the scattering of $(L=0,S=0,T=1)$ nn, np, and pp pairs
as well as $(L=0,S=1,T=0)$ np pairs,
describing the large spatial overlap of the  nucleons' wave function in a relative $L=0$ state.
In the $jj$ coupling scheme the spin-orbit splitting $v_{\ell s}$
increases the energy required to form the ($L=0,S=1,T=0)$ state,
thus favoring a ($J=1, T=0$)  {\it quasi-deuteron} configuration. 
It then seems of interest to consider a more realistic case,
namely that of a single-$j$ shell that incorporates the $jj$ coupling scheme,
more appropriate in heavier nuclei.
The difference between these simple $LS$ and $jj$ models
has been discussed in terms of the BCS approximation~\cite{Bes00}. 

Our approach to study this problem
is to use the shell-model code OXBASH~\cite{Brown88}
with an effective two-body force of the form
\begin{equation}
\hat V(g,x)=-xg\hat V_{J=0,T=1}-(1-x)g\hat V_{J=1,T=0},
\label{e_interaction}
\end{equation}
with 
\begin{equation}
\hat V_{J,T}=
{\frac 1 2}(a^\dag_{jt}\times a^\dag_{jt})^{(J,T)}\cdot(\tilde a_{jt}\times\tilde a_{jt})^{(J,T)},
\label{e_defint}
\end{equation}
where $a^\dag_{jm_jtm_t}$ creates a nucleon
with angular momentum $j$ and projection $m_j$,
isospin $t=\frac12$ and projection $m_t$,
and with $\tilde a_{jm_jtm_t}=(-)^{j+m_j+t+m_t}a_{j-m_jt-m_t}$.
The notation $\times$ implies the coupling to angular momentum $J$ and isospin $T$,
and the dot $\cdot$ denotes a scalar product in angular momentum and isospin.
The Hamiltonian~(\ref{e_interaction}) models the mixture of the two types of competing pairing interactions by the parameter $x$,
with $x=0$ corresponding to the isoscalar and $x=1$ to the isovector limits respectively.
The sign convention in Eq.~(\ref{e_interaction}) is such
that $g$ is positive for an attractive interaction $\hat V(g,x)$.
We consider two spin-orbit partners $f_{7/2}$ and $f_{5/2}$
and study the low-lying spectra obtained as a function of the splitting $v_{\ell s}$.
In the limit $v_{\ell s}=0$ we recover the results of the $LS$ coupling scheme. 
The results for $v_{\ell s} \gg\langle\hat V\rangle$ 
agree with those obtained for a single $f_{7/2}$ level only.

\begin{figure}
\includegraphics[width=0.4\textwidth, angle=90]{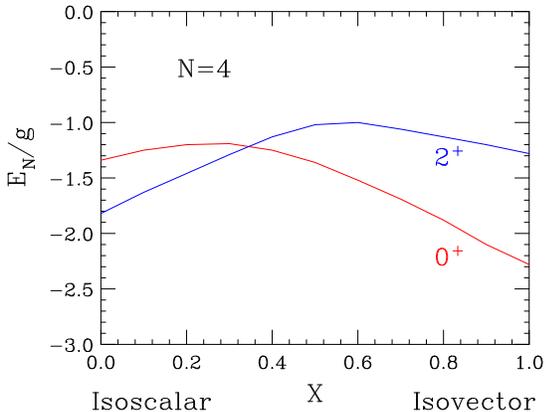} 
\caption{Energies (in units of the pairing strength $g$) of the lowest two $T=0$ states
for $N=4$ particles in an $f_{7/2}$ shell
as a function of the relative mixture $x$ of isovector and isoscalar pairing.}
\label{f_f7n4}
\end{figure}

\begin{figure}
\includegraphics[width=0.4\textwidth, angle=90]{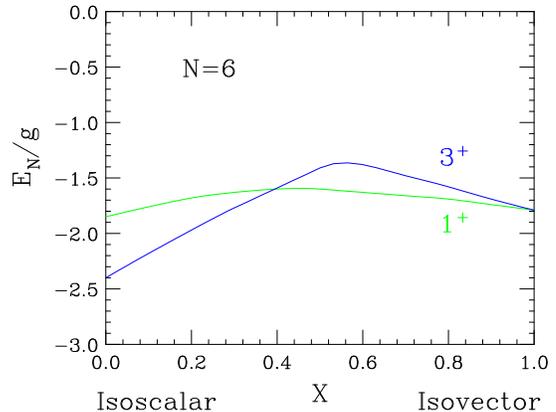} 
\caption{Energies (in units of the pairing strength $g$) of the lowest two $T=0$ states
for $N=6$ particles in the $f_{7/2}$ shell
as a function of the relative mixture $x$ of isovector and isoscalar pairing.}
\label{f_f7n6}
\end{figure}

The intriguing phenomenon that motivated this study is seen in Fig.~\ref{f_f7n4},
showing the evolution of the two lowest states
in the $N=4$ particle system as a function of $x$.
For an appreciable amount of isoscalar pairing ($x\lesssim0.4$) 
the ground state changes from the expected $0^+$ to a $2^+$ state. 
Moreover, as seen in Fig.~\ref{f_f7n6},
the ground state for $N=6$ is a $3^+$ and not $1^+$,
and so on for more particles.
Considering that, for two particles interacting with the force~(\ref{e_interaction}),
$x\sim0$ favors deuteron-like pairing with angular momentum $J=1$, 
it appears that the ground state of the many-particle system
prefers the aligned configuration of the $n=N/2$ pairs,
{\it i.e.} the configuration with $J=n$. 

\begin{figure}
\includegraphics[width=0.4\textwidth,angle=90]{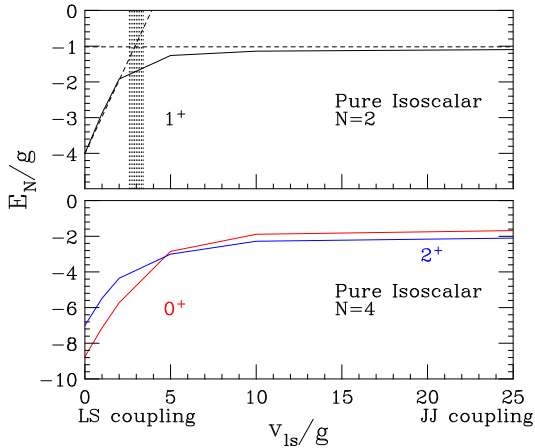} 
\caption{Energies (in units of the pairing strength $g$) of the $T=0$ ground state for $N=2$ particles (top)
and of the lowest two $T=0$ states for $N=4$ particles (bottom)
as a function of the spin-orbit splitting between the $f_{7/2}$ and $f_{5/2}$ orbits
and for pure isoscalar pairing. 
The shaded area in the top panel indicates the critical value of the spin-orbit splitting, $v_{\ell s}^*$,
at which the $jj$ coupling takes on. (See text for details).}
\label{f_fn24}
\end{figure}

We can trace back the change in the properties of the ground state to the spin-orbit splitting.
In Fig.~\ref{f_fn24} we show the results for the $N=4$ system and a pure isoscalar force. 
The energies of the $0^+$ and $2^+$ states
are plotted as a function of the spin-orbit splitting $v_{\ell s}$. 
The two states cross, with a $0^+$ ground state in $LS$ coupling,
which becomes a $2^+$ in $jj$ coupling, as we saw above.

To obtain an estimate of the critical value  $v_{\ell s}^*$ at which the switch occurs,
we consider the case of the $N=2$ system,
also shown in Fig.~\ref{f_fn24}.
Taking the limit of large $j$, to simplify the $LS$-$jj$ re-coupling coefficients,
we have in $jj$ coupling $E_{jj}(1^+)=-g$
and in $LS$ coupling $E_{LS}(1^+)\approx-6g$.

The $^3S_1$ state can be written in terms of the $jj$-coupled wave functions as~\cite{Talmi93}
\begin{equation}
|^3S_1\rangle\approx
\frac{1}{\sqrt{6}} |j^2_>\rangle+
\frac{2}{\sqrt{6}} |j_>j_<\rangle-
\frac{1}{\sqrt{6}} |j^2_<\rangle,
\end{equation}  
from which we can treat perturbatively the effect of the spin-orbit splitting $v_{\ell s}$. 
This gives for an intermediate coupling
\begin{equation}
E_{\rm IC}(1^+)\approx E_{LS}(1^+)+\frac{1}{6} 2 v_{\ell s}+\frac{4}{6}  v_{\ell s}= E_{LS}(1^+)+ v_{\ell s}.
\end{equation}  
The critical value is obtained when the energy above
equals that of the $jj$-coupling limit (dashed lines in Fig.~\ref{f_fn24})
\begin{equation}
E_{jj}(1^+)=E_{LS}(1^+)+ v_{\ell s}^*
\end{equation}  
and we find (in the large-$j$ limit)
\begin{equation}
\frac{v_{\ell s}^*}{g}\approx5.
\end{equation}  
For the particular case of the $f_{7/2}$-$f_{5/2}$ pair (finite-$j$) we find  a  value of $\sim3.5$,
in agreement with the estimate shown in Fig.~\ref{f_fn24} (shaded area).

To shed further light on the properties of the isoscalar condensate discussed above,
we develop in the next section a description based on a mapping
to interacting $p$ bosons of angular momentum $J=1$ and isospin $T=0$.
Based on the results above, and on the fermionic nature of the problem,
we anticipate that the residual interaction between these bosons
favors their aligned coupling.

\section{Isoscalar pairing between fermions in a single $j$ shell}
\label{s_jshell}
Consider a system of $N$ particles, $n$ neutrons and $n$ protons, in a single-$j$ shell,
interacting through an isoscalar pairing interaction
with angular momentum $J=1$ and isospin $T=0$,
corresponding to the $x=0$ limit of Eq.~(\ref{e_interaction}),
$\hat V(g,x=0)=-g\hat V_{10}$.

As discussed in the previous section,
a possible strategy for simplifying the problem
starts from the observation that, by definition of the interaction,
the dominant pair in the two-particle system has $J=1$ and $T=0$.
We attempt to represent a subset of the $2n$-particle eigenstates of this interaction,
including hopefully those at lowest energies,
in terms of a single state $|P\rangle\equiv P^\dag|{\rm o}\rangle$
(with $|{\rm o}\rangle$ the vacuum),
which has $J=1$ and $T=0$,
\begin{equation}
P^\dag_{M_J}\equiv(a^\dag_{jt}\times a^\dag_{jt})^{(J=1,T=0)}_{M_J,M_T=0}.
\label{e_ppair}
\end{equation}
The natural framework to test this idea
is provided by the nucleon-pair shell model (NPSM),
which assumes a basis constructed from nucleon pairs~\cite{Chen93,Chen97,Zhao13un,Fu13}.
In this approximation the full $T=0$ shell-model space
is truncated to one constructed out of $P$ pairs
with basis states $|P^nJ_2\dots J_{n-1}J\rangle$
that are proportional to
\begin{equation}
\left(\cdots\left(\left(P^\dag\times P^\dag\right)^{(J_2)}\times P^\dag\right)^{(J_3)}\times\cdots\times
P^\dag\right)^{(J)}|{\rm o}\rangle.
\label{e_pbasis}
\end{equation}
This $2n$-particle state is characterized
by the set of intermediate angular momenta $\{J_2,\dots,J_{n-1}\}$,
with $J_1=1$ and $J_n=J$, the total angular momentum of the state.
All pairs have $T=0$
and the coupling in isospin need not be considered.
In principle, several intermediate couplings $\{J_2,\dots,J_{n-1}\}$
are possible for a given total angular momentum $J$.
Such is the case for arbitrary pairs but not for $P$ pairs
since the number of independent states
with angular momentum $J$ constructed out of $n$ $P$ pairs
cannot exceed the corresponding number constructed out of $n$ $p$ bosons,
which is 1 if $n-J$ is non-negative and even, and 0 otherwise.
We conclude therefore that, for a given $J$,
at most one state $|P^nJ_2\dots J_{n-1}J\rangle$ exists,
for which the intermediate angular momenta can be chosen as
\begin{equation}
\left\{\begin{array}{lll}
J_i= i\bmod 2,&&1\leq i\leq n-J,\\
J_i=i-n+J,&&n-J\leq i\leq n,\\
\end{array}\right.
\label{e_icoupling}
\end{equation}
where it is implicitly assumed (as will be from now on)
that $n-J$ is non-negative even.
We denote normalized states as $|P^nJ\rangle$,
tacitly assuming the intermediate coupling convention~(\ref{e_icoupling}).
In this convention the paired and spin-aligned states of particular interest here
correspond to the choice
\begin{equation}
\begin{array}{lllll}
\mbox{\rm paired}&:&J_i= i\bmod 2,&&1\leq i\leq n,\\
\mbox{\rm spin-aligned}&:&J_i=i,&&1\leq i\leq n.\\
\end{array}
\label{e_paiali}
\end{equation}

As long as $n\leq(2j+1)/2$ all states~(\ref{e_icoupling}) exist.
This is no longer necessarily true if the shell is more than half filled,
in which case it is advantageous to reconsider the problem in terms of holes.
We then construct basis states $|\tilde P^{2j+1-n}J_2\dots J_{n-1}J\rangle$
that are proportional to
\begin{equation}
\left(\cdots\left(\left(\tilde P\times\tilde P\right)^{(J_2)}\times\tilde P\right)^{(J_3)}\times\cdots\times
\tilde P\right)^{(J)}|\tilde{\rm o}\rangle,
\label{e_hbasis}
\end{equation}
where $|\tilde{\rm o}\rangle$ represents a full shell
and $\tilde P$ annihilates a $P$ pair,
\begin{equation}
\tilde P_{M_J}\equiv(\tilde a_{jt}\times\tilde a_{jt})^{(J=1,T=0)}_{M_J,M_T=0}.
\label{e_hpair}
\end{equation}
The angular momentum and anti-symmetry considerations
concerning the states~(\ref{e_pbasis}) and~(\ref{e_hbasis}) are the same,
and consequently the latter lead to the same allowed basis states~(\ref{e_icoupling})
with $n$ replaced by $\bar n\equiv2j+1-n$.
We denote such states as $|\tilde P^{\bar n}J\rangle$. 

In general, $|P^nJ\rangle$ and $|\tilde P^{\bar n}J\rangle$
are {\em not} the same state,
\begin{equation}
|P^nJ\rangle\neq|\tilde P^{\bar n}J\rangle,
\label{e_phuneq}
\end{equation}
and it is possible
that the state on the left-hand side exists
while the one on the right-hand side does not (or {\it vice versa}).
Only if the shell-model state with a given $J$ and $T=0$ is unique,
do the particle and hole representations become equivalent,
as is the case, for example, for the states
\begin{equation}
|P^{2j+1}J=0\rangle=|\tilde {\rm o}\rangle,
\quad
|P^{2j}J=1\rangle=|\tilde PJ=1\rangle.
\label{e_pheq}
\end{equation}
The choice $|P^nJ\rangle$ if $n\leq(2j+1)/2$
and $|\tilde P^{\bar n}J\rangle$ if $n\geq(2j+1)/2$,
apart from being computationally simpler,
gives the best approximation of shell-model states in terms of $P$ pairs.

The summary of the above discussion is that
the truncated shell-model basis constructed out of $P$ pairs
is spanned by the states $|P^nJ\rangle$ if $n\leq(2j+1)/2$
and by the states $|\tilde P^{\bar n}J\rangle$ if $n\geq(2j+1)/2$.
These basis states exist
(provided $n-J$ or $\bar n-J$ is non-negative even)
and are unique for a given $n$ and $J$,
so that no additional labels are needed.
Therefore, in the $P$-pair approximation of the NPSM,
the correlation energy due to isoscalar pairing
in the state with $n$ neutrons and $n$ protons,
coupled to total angular momentum $J$ and isospin $T=0$,
is\footnote{We reserve the notation $E_{\rm f}(n,J)$ and $E_{\rm f}(\bar n,J)$
({\it i.e.}, expressions without tilde)
for the exact correlation energy of the yrast state with angular momentum $J$
calculated in the full shell-model space.}
\begin{equation}
\tilde E_{\rm f}(n,J)\equiv
\langle P^nJ|-g\hat V_{10}|P^nJ\rangle,
\label{e_etfp}
\end{equation}
for $n\leq(2j+1)/2$, and by
\begin{equation}
\tilde E_{\rm f}(\bar n,J)\equiv
\langle\tilde P^{\bar n}J|-g\hat V_{10}|\tilde P^{\bar n}J\rangle,
\label{e_etfh}
\end{equation}
for $n\geq(2j+1)/2$.
The computation of the matrix elements of an arbitrary interaction
between nucleon-pair states
is possible with the recurrence relation devised by Chen~\cite{Chen97}.
In the general formulation of the NPSM
care should be taken of the over-completeness and non-orthogonality of the pair basis.
This is not an issue in the present application
since basis states are unique for a given $n$ and $J$.
It should be stressed that Eqs.~(\ref{e_etfp}) and~(\ref{e_etfh})
yield an approximation to the exact isoscalar-pairing correlation energy.

The energy $E_{\rm f}(n,J)$ of a particle state
is calculated with respect to the vacuum $|{\rm o}\rangle$
while that of a hole state, $E_{\rm f}(\bar n,J)$,
is with respect to the full shell $|\tilde {\rm o}\rangle$.
The particle-hole transformation gives a relation between both quantities,
which is exact in the full shell-model space.
For our particular case of isoscalar pairing this relation is
\begin{equation} 
E_{\rm f}(n,J)=
-\frac{3(2j+1-2\bar n)}{2j+1}g+E_{\rm f}(\bar n,J).
\label{e_eph}
\end{equation}
We use the same equation
to relate the approximate energies $\tilde E_{\rm f}(n,J)$ and $\tilde E_{\rm f}(\bar n,J)$.
In the following absolute energies
are quoted with respect to the vacuum $|{\rm o}\rangle$.
For a particle state they are obtained directly
while for a hole state they follow from Eq.~(\ref{e_eph}).

A further approximation is to replace
the $P$ pairs by $p$ bosons,
with single-boson energies and boson-boson interactions
derived from the two-particle and four-particle systems, respectively.

With use of the OAI mapping~\cite{Otsuka78}
a $p$-boson Hamiltonian $\hat H_{\rm b}$ is obtained,
which can be written as
\begin{equation}
\hat H_{\rm b}=
\epsilon_p\,p^\dag\cdot\tilde p+
{\frac 1 2}\sum_{\lambda=0,2}v^{\rm b}_\lambda
(p^\dag\times p^\dag)^{(\lambda)}\cdot(\tilde p\times\tilde p)^{(\lambda)},
\label{e_hamb}
\end{equation}
where $\epsilon_p$ is the $p$-boson energy
and $v^{\rm b}_\lambda$ are the two-body interaction matrix elements
{\em between the $p$ bosons}.
The definition of the adjoint operator $\tilde p_m\equiv(-)^{1-m}p_{-m}$
ensures that $\tilde p_m$ is an annihilation operator
with transformation properties under rotations that are the same
as those for the creation operator $p_m^\dag$~\cite{Iachello06}.
With the above definitions we have
that $p^\dag\cdot\tilde p=\sum_m p_m^\dag p_m$
is the number operator $\hat n_p$.

The single-boson energy is
\begin{equation}
\epsilon_p\equiv
\langle p|\hat H_{\rm b}|p\rangle\doteq
\langle P|-g\hat V_{10}|P\rangle=-g.
\label{e_ebos}
\end{equation}
where the notation $\doteq$ is used to indicate
that the equality holds by virtue of the mapping procedure.
The two-body boson matrix elements with $\lambda=0,2$ are
\begin{align}
v^{\rm b}_\lambda&\equiv
\langle p^2\lambda|\hat H_{\rm b}|p^2\lambda\rangle-2\epsilon_p
\nonumber\\&\doteq
-g\left(\langle j^4[10,10]\lambda0
|\hat V_{10}|
j^4[10,10]\lambda0\rangle-2\right),
\label{e_hmat4}
\end{align}
where the bra and ket represent normalized, anti-symmetric two-pair states,
\begin{equation}
|j^4[J_1T_1,J_2T_2]JT\rangle
\propto{\cal A}|j^2(J_1T_1)j^2(J_2T_2)JT\rangle.
\label{e_ket4}
\end{equation}
The notation in square brackets $[J_1T_1,J_2T_2]$ implies that the state~(\ref{e_ket4})
is constructed from a parent state
with intermediate angular momenta and isospins $J_1T_1$ and $J_2T_2$.
The anti-symmetrized states $|j^4[J_1T_1,J_2T_2]JT\rangle$
can be expanded in terms of the two-pair states $|j^2(J_1T_1)j^2(J_2T_2)JT\rangle$
by means of four-to-two-particle coefficients of fractional parentage (CFPs)~\cite{Talmi93},
\begin{equation}
[j^2(J_aT_a)j^2(J_bT_b)JT|\}j^4[J_1T_1,J_2T_2]JT],
\label{e_cfp}
\end{equation}
which are known in closed form.

From the general expression for the matrix element~(\ref{e_hmat4})
the following results are obtained:
\begin{align}
v^{\rm b}_0/g={}&
-6[j^2(10)j^2(10)00|\}j^4[10,10]00]^2+2,
\nonumber\\
v^{\rm b}_2/g={}&
-6[j^2(10)j^2(10)20|\}j^4[10,10]20]^2
\nonumber\\&-
6[j^2(30)j^2(10)20|\}j^4[10,10]20]^2+2,
\end{align} 
which, with the help of
\begin{align}
&[j^2(10)j^2(10)00|\}j^4[10,10]00]^2=
\frac{2j^3-2j+3}{3j(j+1)(2j+1)},
\nonumber\\
&[j^2(10)j^2(10)20|\}j^4[10,10]20]^2=
\frac{10j^3+9j^2-j-3}{15j(j+1)(2j+1)},
\nonumber\\
&[j^2(30)j^2(10)20|\}j^4[10,10]20]^2
\nonumber\\&\qquad=
\frac{9(j-1)(j+2)(2j+3)}{10j(j+1)(2j+1)(5j^2+7j+3)},
\end{align} 
lead to the following expressions for the $p$-boson matrix elements:
\begin{align}
v^{\rm b}_0={}&
\frac{6(j^2+j-1)}{j(j+1)(2j+1)}g
\stackrel{j\rightarrow\infty}{\longrightarrow}
\left[\frac{3}{j}+
{\cal O}\left(\frac{1}{j^2}\right)\right]g,
\nonumber\\
v^{\rm b}_2={}&
\frac{3(4j^4+6j^3+j^2+7j+12)}{j(j+1)(2j+1)(5j^2+7j+3)}g
\nonumber\\&
\stackrel{j\rightarrow\infty}{\longrightarrow}
\left[\frac{6}{5j}+
{\cal O}\left(\frac{1}{j^2}\right)\right]g.
\label{e_vbos}
\end{align}
As anticipated, for an attractive isoscalar pairing interaction
the boson-boson matrix elements are repulsive.
This is a finite-space effect, due to the Pauli principle,
since the matrix elements vanish in the large-$j$ limit.
A difference between the $\lambda=0$ and $\lambda=2$ matrix elements
also arises due to Pauli effects,
and it is seen that $v^{\rm b}_2$ is less repulsive.
This favors the spin-aligned ground state,
not only for two but also for more bosons
as a result of the following argument.

Since a system of $n$ interacting identical $p$ bosons is solvable
by virtue of a ${\rm U}(3)\supset{\rm SO}(3)$ dynamical symmetry~\cite{Iachello06},
the eigenvalues of the Hamiltonian~(\ref{e_hamb})
are known in closed form,
\begin{align}
E_{\rm b}(n,J)={}&
n\epsilon_p+
\frac{n(n+1)-J(J+1)}{6}v^{\rm b}_0
\nonumber\\&+
\frac{2n(n-2)+J(J+1)}{6}v^{\rm b}_2,
\label{e_enebos}
\end{align}
where the allowed angular momenta are $J=n,n-2,\dots,1$ or 0.
The only possible ground states of a $p$-boson system
are either paired or spin-aligned~\cite{Law98,Isacker07}.
The paired state has $J=0$ or $J=1$ with energies
\begin{align}
E_{\rm b}(n,J=0)&=
n\epsilon_p+
\frac{n(n+1)}{6}v^{\rm b}_0+
\frac{n(n-2)}{3}v^{\rm b}_2,
\label{e_enepaired}\\
E_{\rm b}(n,J=1)&=
n\epsilon_p+
\frac{(n-1)(n+2)}{6}v^{\rm b}_0+
\frac{(n-1)^2}{3}v^{\rm b}_2,
\nonumber
\end{align}
depending on whether $n$ is even or odd, respectively.
The spin-aligned state has $J=n$ with energy
\begin{equation}
E_{\rm b}(n,J=n)=
n\epsilon_p+
\frac{n(n-1)}{2}v^{\rm b}_2.
\label{e_enealigned}
\end{equation}
The breaking of the rotational invariance in {\it gauge space}~\cite{broglia}
leads to the emergence of isoscalar pairing rotational bands,
as seen in the quadratic dependence of the energies
as a function of the number of pairs $n$,  Eqs.~(\ref{e_enepaired}) and (\ref{e_enealigned}).

The difference in energy
between the paired and the spin-aligned states
can be written as
\begin{align}
\Delta_{\rm b}(n)&=
\frac{(n-n_2)(n+1+n_2)}{6}(v^{\rm b}_0-v^{\rm b}_2)
\nonumber\\&\approx
g\frac{3(n-n_2)(n+1+n_2)}{10j},
\end{align}
where $n_2$ is 0 for even $n$ and 1 for odd $n$,
 $n_2\equiv n \bmod 2$.
This shows that for all $n$ the difference in energy
between the paired and the spin-aligned states
is positive for an attractive pairing interaction,
that is, the spin-aligned configuration is the ground state.

\begin{table}
\centering
\caption{\label{t_shellj}
Exact energies $E_{\rm f}(n,J)$ of paired ($J=0$ or 1) and aligned ($J=n$) states with $T=0$
of a system of $n$ neutrons and $n$ protons in a single-$j$ shell
interacting through an isoscalar pairing force,
in units of the strength $g$,
and the corresponding energies $\tilde E_{\rm f}(n,J)$ and $E_{\rm b}(n,J)$
obtained in the $P$-pair and $p$-boson approximations.
A dash --- means that a $P$-pair state does not exist
while the absence of an entry indicates
that the numerical result could not be obtained.}
\smallskip
\begin{tabular}{ccccccccccc}
\hline\hline
$E(n,J)$&&$j=7/2$&~&$j=9/2$&~&$j=11/2$&~&$j=13/2$&~&$j=15/2$\\
\hline
$E_{\rm f}(2,0)$&&$-1.298$ &&$-1.424$ &&$-1.514$ &&$-1.580$ &&$-1.631$\\
$\tilde E_{\rm f}(2,0)$&&$-1.298$ &&$-1.424$ &&$-1.514$ &&$-1.580$ &&$-1.631$\\
 $E_{\rm b}(2,0)$&&$-1.298$ &&$-1.424$ &&$-1.514$ &&$-1.580$ &&$-1.631$\\
$E_{\rm f}(2,2)$&&$-1.793$ &&$-1.825$ &&$-1.847$ &&$-1.865$ &&$-1.879$\\
$\tilde E_{\rm f}(2,2)$&&$-1.757$ &&$-1.799$ &&$-1.828$ &&$-1.850$ &&$-1.866$\\
$E_{\rm b}(2,2)$&&$-1.757$ &&$-1.799$ &&$-1.828$ &&$-1.850$ &&$-1.866$\\
\hline
$E_{\rm f}(3,1) $&&$-1.793$ &&$-1.953$ &&$-2.086$ &&$-2.192$ &&$-2.277$\\
$\tilde E_{\rm f}(3,1)$&&$-1.636$ &&$-1.848$ &&$-2.010$ &&$-2.135$ &&$-2.233$\\
$E_{\rm b}(3,1)$&&$-1.505$ &&$-1.772$ &&$-1.961$ &&$-2.100$ &&$-2.207$\\
$E_{\rm f}(3,3) $&&$-2.365$ &&$-2.466$ &&$-2.537$ &&$-2.591$ &&$-2.634$\\
$\tilde E_{\rm f}(3,3)$&&$-2.279$ &&$-2.403$ &&$-2.488$ &&$-2.552$ &&$-2.601$\\
$E_{\rm b}(3,3)$&&$-2.271$ &&$-2.397$ &&$-2.484$ &&$-2.549$ &&$-2.599$\\
\hline
$E_{\rm f}(4,0) $&&$-2.080$ &&$-2.251$ &&$-2.424$ &&$         $ &&$         $\\
$\tilde E_{\rm f}(4,0)$&&$-1.628$ &&$-1.887$ &&$-2.141$ &&$-2.353$ &&$-2.526$\\
$E_{\rm b}(4,0)$&&$-1.010$ &&$-1.545$ &&$-1.921$ &&$-2.200$ &&$-2.413$\\
$E_{\rm f}(4,4) $&&$-2.767$ &&$-2.925$ &&$         $ &&$         $ &&$         $\\
$\tilde E_{\rm f}(4,4)$&&$-2.577$ &&$-2.818$ &&$-2.985$ &&$-3.110$ &&$-3.207$\\
$E_{\rm b}(4,4)$&&$-2.541$ &&$-2.794$ &&$-2.968$ &&$-3.098$ &&$-3.198$\\
\hline
$E_{\rm f}(5,1) $&&$-2.543$ &&$         $ &&$         $ &&$         $ &&$         $\\
$\tilde E_{\rm f}(5,1)$&&$-2.386$ &&$-1.975$ &&$-2.284$ &&$-2.566$ &&$-2.806$\\
$E_{\rm b}(5,1)$&&$-2.255$ &&$-1.241$ &&$-1.815$ &&$-2.239$ &&$-2.564$\\
$E_{\rm f}(5,5) $&&--- &&$         $ &&$         $ &&$         $ &&$         $\\
$\tilde E_{\rm f}(5,5)$&&--- &&$-3.052$ &&$-3.324$ &&$-3.528$ &&$-3.687$\\
$E_{\rm b}(5,5)$&&--- &&$-2.990$ &&$-3.281$ &&$-3.496$ &&$-3.663$\\
\hline
$E_{\rm f}(6,0) $&&$-2.798$ &&$-2.851$ &&$         $ &&$         $ &&$         $\\
$\tilde E_{\rm f}(6,0)$&&$-2.798$ &&$-2.487$ &&$-2.206$ &&$-2.539$ &&$-2.847$\\
$E_{\rm b}(6,0)$&&$-2.798$ &&$-2.145$ &&$-1.222$ &&$-1.858$ &&$-2.347$\\
$E_{\rm f}(6,6) $&&--- &&--- &&$         $ &&$         $ &&$         $\\
$\tilde E_{\rm f}(6,6)$&&--- &&--- &&$-3.511$ &&$-3.810$ &&$-4.045$\\
$E_{\rm b}(6,6)$&&--- &&--- &&$-3.421$ &&$-3.744$ &&$-3.994$\\
\hline
$E_{\rm f}(7,1) $&&$-3.250$ &&$-3.153$ &&$         $ &&$         $ &&$         $\\
$\tilde E_{\rm f}(7,1)$&&$-3.250$ &&$-3.048$ &&$-2.784$ &&$-2.534$ &&$-2.892$\\
$E_{\rm b}(7,1)$&&$-3.250$ &&$-2.972$ &&$-2.315$ &&$-1.417$ &&$-2.074$\\
$E_{\rm f}(7,7) $&&--- &&--- &&--- &&$     $ &&$         $\\
$\tilde E_{\rm f}(7,7)$&&--- &&--- &&--- &&$-3.962$ &&$-4.284$\\
$E_{\rm b}(7,7)$&&--- &&--- &&--- &&$-3.842$ &&$-4.192$\\
\hline
$E_{\rm f}(8,0) $&&$-3.000$ &&$-3.224$ &&$-3.424$ &&$         $ &&$         $\\
$\tilde E_{\rm f}(8,0)$&&$-3.000$ &&$-3.224$ &&$-3.141$ &&$-2.968$ &&$-2.776$\\
$E_{\rm b}(8,0)$&&$-3.000$ &&$-3.224$ &&$-2.921$ &&$-2.287$ &&$-1.431$\\
$E_{\rm f}(8,8) $&&--- &&--- &&--- &&--- &&$         $\\
$\tilde E_{\rm f}(8,8)$&&--- &&--- &&--- &&--- &&$-4.408$\\
$E_{\rm b}(8,8)$&&--- &&--- &&--- &&--- &&$-4.256$\\
\hline\hline
\end{tabular}
\end{table}
We recall that the preceding results,
valid for an isoscalar pairing interaction in a single-$j$ shell,
are derived under the following simplifying assumptions:
\begin{enumerate}
\item
The full shell-model space is truncated
to one constructed out of $P$ pairs.
The expectation value of the isoscalar pairing Hamiltonian $-g\hat V_{10}$
in the (unique) $P$-pair state
takes fully account of the Pauli principle
and leads to the approximate correlation energy $\tilde E_{\rm f}(n,J)$.
\item
The fermionic Hilbert space constructed out of $P$ pairs
is mapped onto a corresponding bosonic Hilbert space constructed out of $p$ bosons.
The mapping of the Hamiltonian is carried out in the two- and four-nucleon spaces
and leads to a boson Hamiltonian with up to two-body interactions.
\item
The boson Hamiltonian is used
to calculate the energies $E_{\rm b}(n,J)$ of $n$-boson states.
\end{enumerate}

To gauge the adequacy of the different approximations,
we show in Table~\ref{t_shellj} the exact energies $E_{\rm f}(n,J)$
(wherever they can be calculated)
and the corresponding approximations $\tilde E_{\rm f}(n,J)$ and $E_{\rm b}(n,J)$
for $7/2\leq j\leq15/2$.
Several comments are in order.
First of all, we observe the identity
\begin{equation} 
\tilde E_{\rm f}(n=2,J)=E_{\rm b}(n=2,J),
\label{e_n2ident}
\end{equation}
that is, the $P$-pair spectrum of the four-particle system
coincides with that obtained for two $p$ bosons.
This is a generic property of the mapping
and follows from the fact that up to two-body interactions
between the bosons are considered.
In fact, if up to $q$-body interactions are considered,
the identity~(\ref{e_n2ident}) remains valid up to the $n=q$.
Secondly, we observe the identity
\begin{equation} 
E_{\rm f}(n=2,J=0)=\tilde E_{\rm f}(n=2,J=0).
\label{e_n2j0dent}
\end{equation}
This is not a generic property 
but is valid for the isoscalar pairing interaction,
for which $|P^2J=0\rangle$ decouples from the rest of the shell-model space.
This property of the isoscalar pairing interaction
was already pointed out by Fu {\it et al.}~\cite{Fu14}
on the basis of analytic expressions for four-nucleon overlaps.
Furthermore, we observe from Table~\ref{t_shellj} the following hierarchy:
\begin{equation} 
E_{\rm f}(n,J)\leq\tilde E_{\rm f}(n,J)\leq E_{\rm b}(n,J),
\label{e_inequal}
\end{equation}
valid for any $j$, $n$, and $J$.
The first inequality results from the fact that
the lowest eigenvalue of any Hamiltonian in a certain Hilbert space
is lower than the lowest eigenvalue of the same Hamiltonian in a truncated subspace.
We remark that an equality $E_{\rm f}(n=4,J)=\tilde E_{\rm f}(n=4,J)$ can be obtained
by constructing {\em effective} operators in the truncated space,
which is not done in the present application.
The second inequality in Eq.~(\ref{e_inequal})
is a consequence of performing the mapping in the four-particle systems
with an unnormalized ({\it i.e.}, not an effective) Hamiltonian.
For a variety of bosonic systems ($p$, $sd$, $sdp$, etc.)
we have consistently found that the boson Hamiltonian,
as it is derived here from the four-particle system,
gives an upper limit for the fermionic interaction energy of the $n$-particle system.

It is seen from Table~\ref{t_shellj} that the quality of the approximation
varies with $j$, $n$, and $J$.
Two effects are rather obvious:
the approximation becomes (i) better with increasing $j$
and (ii) worse with increasing $n$ [as long as $n\leq(2j+1)/2$].
These effects result from the increasing importance
of Pauli corrections that are neglected
({\it i.e.}, beyond two-body interactions between the bosons).
A more subtle effect is the dependence on $J$.
It is seen that the approximation for the aligned state $J=n$ is adequate,
even close to mid shell, $n\approx(2j+1)/2$, and for low $j$.
On the other hand, it is often rather poor for the paired state with $J=0$ or 1.
It can be conjectured that this is a generic property
of phonon approximations in fermionic systems:
while such descriptions are good for high-angular-momentum states,
they become highly anharmonic at low angular momenta.

Despite the varying quality of the boson approximation,
depending on $j$, $n$, and $J$,
the overall conclusion is that
the predicted feature of the lower energy of the aligned state
as compared to the paired state
is confirmed by the exact fermion calculation.

\section{Conclusion}
\label{s_conc}
We have considered some intriguing properties of a $T=0$ isoscalar condensate in single $j$-shell,
in particular with regards to its angular momentum coupling. 
We developed a description based on a mapping of the shell model
to interacting $p$ bosons of angular momentum $J=1$ and isospin $T=0$,
providing for the first time approximate analytic formulae
for the energies of the lowest states.
Our results show that, due to the Pauli principle,
the residual interaction between these bosons favors 
({\it a priori } unexpected) the aligned configuration of  $n=N/2$ {\sl quasi-deuteron} pairs,
{\it i.e.} that with $J=n$.

While we realize this limit may not be applicable to real nuclei,
these states might exist close to the paired ground states
in specific regions close to the $N=Z$ line. 
In fact, it was shown recently in Refs.~\cite{Gezerlis11,Bulthuis16},
using a phenomenological Hamiltonian within the framework of the Hartree-Fock-Bogoliubov theory,
that the spin-triplet phase is favored over the spin-singlet one in the mass region $ A\approx130$  with $Z \approx64$
(for example $^{132}$Gd).
This is found to depend on the occupation of specific low-$j$ orbitals
near the Fermi energy for which the spin-orbit splitting is small.
More interestingly perhaps,
it is envisioned that these condensates might be realized by tunable spin-orbit coupling in ultracold atomic traps~\cite{Lin11},
whereby the control parameter $x$ in Eq.~(\ref{e_interaction})
could be adjusted to  drive the system from diamagnetic to magnetic.

A full extension of the present formalism
including the effect of the spin-orbit splitting
will be the subject of a future publication.

\section{Ackowledgements}
This work was supported in part by the FUSTIPEN
(French-U.S. Theory Institute for Physics with Exotic Nuclei)
under U.S. DOE grant No.~ DE-FG02-10ER41700, and by the U.S. DOE contract
No.~DE-AC02-05CH11231 (LBNL).

\end{document}